\documentclass[onecolumn,11pt,superscriptaddress]{revtex4-1}
\usepackage{bm}        % needed for bold math
\usepackage{graphicx}  % needed for figures
\usepackage{verbatim}  % useful for program listings
\usepackage{color}     % use if color is used in text
\usepackage{multirow}  %
\usepackage{dcolumn}
\usepackage{lipsum}
\usepackage{subfigure}
\usepackage{float}
\usepackage{natbib}
\usepackage{longtable}
\usepackage{rotating}

\begin{document}
%\begin{CJK*}{GBK}{}
\title{Tungsten spectroscopy in the EUV observed in SH-HtscEBIT}
\date{\today}

\author{Wenxian Li}
\affiliation{The Key lab of Applied Ion Beam Physics, Ministry of Education, China}
\affiliation{Shanghai EBIT laboratory, Institute of Modern physics, Fudan University, Shanghai, China}

\author{Zhan Shi}
\affiliation{Division of Mathematical Physics, Department of Physics, Lund University, Sweden}

\author{Yang Yang}
\affiliation{The Key lab of Applied Ion Beam Physics, Ministry of Education, China}
\affiliation{Shanghai EBIT laboratory, Institute of Modern physics, Fudan University, Shanghai, China}

\author{Jun Xiao}
\affiliation{The Key lab of Applied Ion Beam Physics, Ministry of Education, China}
\affiliation{Shanghai EBIT laboratory, Institute of Modern physics, Fudan University, Shanghai, China}

\author{Tomas Brage}
\affiliation{Division of Mathematical Physics, Department of Physics, Lund University, Sweden}

\author{Roger Hutton}\email{rhutton@fudan.edu.cn}
\affiliation{The Key lab of Applied Ion Beam Physics, Ministry of Education, China}
\affiliation{Shanghai EBIT laboratory, Institute of Modern physics, Fudan University, Shanghai, China}

\author{Yaming Zou}
\affiliation{The Key lab of Applied Ion Beam Physics, Ministry of Education, China}
\affiliation{Shanghai EBIT laboratory, Institute of Modern physics, Fudan University, Shanghai, China}

\begin{abstract}
We have recorded extreme ultraviolet spectra from $\mathrm{W^{11+}}$ to $\mathrm{W^{15+}}$ ions using a new flat field spectrometer installed at the Shanghai high temperature superconducting electron beam ion trap. The spectra were recorded at beam energies ranging between 200 eV and 400 eV and showed spectral lines/transition arrays in the 170~-~260~\AA{} region. The charge states and spectra transitions were identified by comparison with calculations using a detailed relativistic configuration interaction method and collisional-radiative model, both incorporated in the Flexible Atomic Code. Atomic structure calculations showed that the dominant emission arises from $5d$ $\rightarrow$ $5p$ and $5p$ $\rightarrow$ $5s$ transitions.
The work also identified the ground-state configuration of $W^{13+}$ as $4f^{13}5s^2$ both theoretically and experimentally.

\end{abstract}

\maketitle

\section{Introduction}
There has been a strong interest in tungsten spectroscopy due to the potential use of this element as a plasma facing material in the the International Thermonuclear Experimental Reactor (ITER) tokamak~\cite{Putterich2005,Skinner2008}, especially in the divertor region~\cite{Philipps2006}. Spectra diagnostics of the ITER divertor region will therefore require a large amount of, as yet, unavailable tungsten atomic data. The ITER divertor soft x-ray spectrometer will operate in the 150~-~400~\AA{} region~\cite{Robin2014} where only 108 lines are known~\cite{NIST}. Of these 102 lines originate from $\mathrm{W^{4+}}$, $\mathrm{W^{5+}}$ and $\mathrm{W^{6+}}$. One line is identified as from $\mathrm{W^{51+}}$ which is too highly ionized to be observed in the divertor region. The plasma temperature in the divertor region of ITER is expected to be in the region of a few to a few hundred eV which will result in tungsten in charges states of up to around 28+. The remaining 5 tungsten lines in this wavelength region are reported to be from $\mathrm{W^{13+}}$, but might be mis-identified ~\cite{Trabert1986}. In contrast to this, the diagnostics of the core plasma region has a stronger support, since considerably more spectroscopic work, both experimental and theoretical, has been reported for the important charge states, (see the review by Kramida~\cite{Kramida2011}).

In previous papers we have investigated visible M1 transitions in $\mathrm{W^{13+}}$~\cite{Zhao2014}, $\mathrm{W^{25+}}$~\cite{Li2015}, $\mathrm{W^{26+}}$~\cite{Fei2014}, $\mathrm{W^{27+}}$~\cite{Fei2012} (with an extension to the silver iso-electronic sequence~\cite{ruifeng2014,Grumer2014}) and $\mathrm{W^{28+}}$~\cite{Qiu2014}. In the present work we report on a study of tungsten ions in the wavelength region of interest to ITER divertor diagnostics, i.e. 200~-~400 \AA{}. Using the Shanghai high temperature super conducting Electron Beam Ion Trap (SH-HtscEBIT), which was designed for low electron beam energy operation, we observed spectra from $\mathrm{W^{11+}}$ through to $\mathrm{W^{15+}}$. The spectra were recorded using a recently developed high resolution flat field spectrometer~\cite{Zhan2014}. From calculations of the relevant atomic structure and also collisional radiative model, using the FAC code~\cite{Gu2008}, we have identified lines and spectral features as originating from $5d$ $\rightarrow$ $5p$ and $5p$ $\rightarrow$ $5s$ transitions. We will also discuss the interesting case of $\mathrm{W^{13+}}$ in more detail, due to the controversy over the classification of the ground state which has existed since the calculations by Curtis and Ellis in 1980~\cite{Curtis1980}.

\section{Experimental method}
The experiment was carried out using the SH-HtscEBIT, which is described elsewhere~\cite{Xiao2014}. This EBIT is capable of operating in the range of electron beam energies between 30 and 4000~eV. The magnetic field, which is created by liquid nitrogen temperature superconducting coils, compresses the beam radius to $60~\mu m$. The background vacuum pressure in the trap center is estimated to be lower than $10^{-10}$ torr, which makes it possible to produce tungsten ions mainly through electron collisional ionization with negligible influence from charge exchange. The spectra were recorded by utilizing a high resolution grazing-incidence flat field spectrometer, which covers the range of 10 to 500~\AA{} and reaches a resolving power of above 800~\cite{Zhan2014}. In order to eliminate light from the hot cathode, a 4500 \AA{} thick aluminum foil was used and mounted on the window of the mini ultra high vacuum gate valve between the SH-HtscEBIT and the spectrometer. Due to the aluminum L absorption edge, only wavelengths longer than 171 \AA{} can pass through this foil.
For the present experiment, a Shimazdu varied-line-spacing (VLS) grating (1200 l/mm, part number: 001-0659)~\cite{Kita1983,Harada1980,Harada1999} was used and an Andor CCD camera (model number: DO936N-00W-\#BN)  was placed at different positions to record different wavelength regions. Tungsten ions were obtained by injecting $\mathrm{W(CO)_6}$, a volatile compound with a high vapour pressure at room temperature~\cite{Ken2000}. The experiments were done using electron beam energies ranging from 200 to 400 eV, in steps of about 20 eV, and a beam current of 8.1 mA. The spectrometer was calibrated by several background oxygen and nitrogen lines (see table~\ref{calibration}).
\begin{table*}
\caption{\label{calibration}The calibration lines used in this work, from the background oxygen(O IV~\cite{Edlen1963}, O V~\cite{Bockasten1968}) and nitrogen(N IV~\cite{Moore1971}, N V~\cite{Edlen1934}) lines.}
\begin{tabular}{llccccccccccccccc}
\hline
\hline
Spectra   &&  O V    &&  O V   &&   N V   &&    N IV    &&   O IV    \\
Transition   &&  $\mathrm{^1P_1~-~^1S_0}$    &&  $\mathrm{^3D_{1,2}~-~^3P_1}$   &&   $\mathrm{^2P_{1/2}~-~^2S_{1/2}}$   &&    $\mathrm{^3D_{1,2,3}~-~^3P_2}$    &&   $\mathrm{^2D_{5/2}~-~^2P_{3/2}}$    \\
Wavelength(\AA) && 172.169  &&   192.799   &&   209.303 &&  225.210   &&  238.570  \\
\hline
\hline
\end{tabular}
\end{table*}

\section{Description of the calculation}

\begin{table*}
\caption{\label{model} Configurations included in the CR-model for $\mathrm{W^{11+}}$ to $\mathrm{W^{15+}}$ (see text), together with the resulting number of excited levels. We also give the predicted ground state of each charge state in LS coupling and jj-coupling.}
\tiny
\begin{tabular}{lllllllllllllllllllll}
\hline
\hline
ions:          &&  $\mathrm{W^{11+}}$           && $\mathrm{W^{12+}}$  &&    $\mathrm{W^{13+}}$     &&      $\mathrm{W^{14+}}$  &&     $\mathrm{W^{15+}}$          \\
\hline
\# levels:   && 2538  &&  1769 &&  1196   &&  3896  &&  3655  \\
\hline
                &&   $4f^{14}5s^25p$      &&  $4f^{14}5s^2$       &&   $4f^{14}5s$         &&  $4f^{14}$           &&   $4f^{13}$          \\
                &&   $4f^{14}5s5p^2$      &&  $4f^{14}5s5p$       &&   $4f^{14}5p$         &&  $4f^{13}5s$         &&   $4f^{12}5s$        \\
                &&   $4f^{14}5s5p5d$      &&  $4f^{14}5p^2$       &&   $4f^{14}5d$         &&  $4f^{13}5p$         &&   $4f^{12}5p$        \\
                &&   $4f^{13}5s^{2}5p^2$  &&  $4f^{14}5s5d$       &&   $4f^{13}5s^2$       &&  $4f^{13}5d$         &&   $4f^{12}5d$        \\
         &&   $4f^{13}5s^{2}5p5d$  &&  $4f^{14}5p5d$       &&   $4f^{13}5s5p$       &&  $4f^{12}5s^2$       &&   $4f^{11}5s^2$      \\
                &&   $4f^{13}5s5p^3$      &&  $4f^{14}5d^2$       &&   $4f^{13}5p^2$       &&  $4f^{12}5s5p$       &&   $4f^{11}5s5p$      \\
                &&   $4f^{13}5s5p^25d$    &&  $4f^{13}5s^{2}5p$   &&   $4f^{13}5s5d$       &&  $4f^{12}5p^2$       &&   $4f^{11}5p^2$      \\
                &&   $4f^{12}5s^25p^3$    &&  $4f^{13}5s^{2}5d$   &&   $4f^{12}5s^25p$     &&  $4f^{12}5s5d$       &&   $4f^{11}5s5d$      \\
                &&   $4f^{12}5s^25p^25d$  &&  $4f^{13}5s5p^2$     &&   $4f^{12}5s^25d$     &&  $4f^{11}5s^25p$     &&   $4f^{10}5s^25p$    \\
                &&                        &&  $4f^{13}5s5d^2$     &&   $4f^{12}5s5p^2$     &&  $4f^{11}5s^25d$     &&   $4f^{10}5s^25d$    \\
                &&                        &&  $4f^{13}5s5p5d$     &&   $4f^{11}5s^25p^2$   &&  $4f^{11}5s5p^2$     &&                      \\
                &&                        &&  $4f^{12}5s^{2}5p^2$ &&                       &&  $4f^{10}5s^25p^2$   &&                      \\
                &&                        &&  $4f^{12}5s^{2}5p5d$ &&                       &&                      &&                      \\
                &&                        &&  $4f^{12}5s^{2}5d^2$ &&                       &&                      &&                      \\
\hline
ground     &&     $4f^{13}5s^{2}5p^2~ ^4\mathrm{F_{7/2}}$   &&  $4f^{14}5s^2~^1\mathrm{S}_0$      &&   $4f^{13}5s^2~^2\mathrm{F}_{7/2}$   &&   $4f^{12}5s^2~^3\mathrm{H}_6$    &&   $4f^{11}5s^2~^4\mathrm{I}_{15/2}$   \\
 state: &&  $[(4f^{13})_{7/2}5s^2(5p^2_{1/2})_0]_{7/2}$ && $[4f^{14}5s^2]_0$ &&  $[(4f^{13})_{7/2}5s^2]_{7/2}$  &&  $[[(4f^6_{5/2})_0(4f^6_{7/2})_6]_65s^2]_6$ && $[[(4f^6_{5/2})_0(4f^5_{7/2})_{15/2}]_{15/2}5s^2]_{15/2}$ \\
\hline
\hline
\end{tabular}
\end{table*}

In order to support the identification of the recorded spectra and to predict the charge state distributions we performed calculations using the Flexible Atomic Code, FAC v1.1.1.~\cite{Gu2008}. This is an integrated software package~\cite{Gu2004,Gu2005,Gu2006} producing both structure and scattering data, including e.g. energy levels, radiative transition rates, collisional excitation and ionization by electron impact. FAC also includes a collisional radiative(CR) model to produce synthetic spectra for plasmas under different physical conditions. The atomic structure calculation in FAC is based on a relativistic configuration interaction(RCI) model using independent particle basis wavefunctions, derived from a local central potential. The orbitals are optimized in a self-consistent-field (SCF) iterative procedure. Relativistic effects are included through the Dirac-Coulomb Hamiltonian being used for the optimization. Corrections from Breit interacting and higher order QED effects, e.g.vacuum polarization and self-energy, are included in a final CI-calculation. This approach yields energy levels, radiative and autoionization rates, and collision strength from the final wavefunctions~\cite{Gu2004}. The CR-model has been shown to be a reliable tool for the analysis of optically thin plasmas such as those in an EBIT~\cite{Ralchenko2007}. CR-model can provide information on level populations and spectral line intensities and therefore be used for the generation of synthetic spectra. In this model, we have considered electron-impact excitation and de-excitation, together with radiative transition, but neglected other effects, e.g. charge exchange and radiative recombination, since they are less important in dilute EBIT plasma. For a given excited level, with the normalization condition $\sum_iN_i~=~1$, the population was obtained by solving quasi-stationary-state rate equations $\frac{dN_i}{dt}~=~0$, where

\begin{equation}
\label{rates}
\frac{dN_i}{dt} = \sum_{j \ge i}(A^r_{j \rightarrow i} \cdot N_j)+\sum_{j \le i}(C^e_{j \rightarrow i} \cdot N_j)+\sum_{j \ge i}(C^d_{j \rightarrow i} \cdot N_j)-\sum_{j \le i}(A^r_{i \rightarrow j} \cdot N_i)-\sum_{j \ge i}(C^e_{i \rightarrow j} \cdot N_i)-\sum_{j \le i}(C^d_{i\rightarrow j} \cdot N_i)
\end{equation}
In this equation $A^r_{j \rightarrow i}$, $C^e_{j \rightarrow i}$, $C^d_{j \rightarrow i}$ and $N_j$ are the radiative transition rate from level j to i, the collisional excitation rate from j to i, the collisional de-excitation rate from j to i and the population of level j, respectively.

Generally the accuracy of our calculations depends on the size and completeness of the atomic model used, which is determined by the choice of included configurations. Given the complexity of tungsten ions in the range of charge state of interest here we were forced to consider a limited number of levels in our CR-model. One important consideration was to include all important metastable levels since they have long lifetimes, which in turn also gives high population, even relative to the ground state~\cite{Qiu2014,Zhao2014}.  These properties of the metastable levels can lead to two step excitation and ionisation~\cite{Qiu2014}. In  $\mathrm{W^{6+}}$ to $\mathrm{W^{28+}}$ we will deal with configurations with open $4f$-subshells, which are known to give rise to several metastable levels due to the high angular momenta involved.  To give models of reasonable complexities, we restricted our model to include configurations  created by single and double excitation from 4f, 5s and 5p subshells, as shown in table \ref{model}, where we also give the total number of states included for each ion.
The synthetic spectra were create using CR-models for beam energies of 230, 270, 310, 350 and 390 eV, respectively, for $\mathrm{W^{11+}}$ to $\mathrm{W^{15+}}$ at an electron density of $5 \times 10^{10}~\mathrm{cm}^{-3}$, which is a typical electron density for the SH-HtscEBIT. The calculated lines are convolved with a Gaussian line shape to match the spectrometer resolution of 850.

\section{Results and Discussions}
\begin{figure*}
\centering
\includegraphics[width=0.85\textwidth]{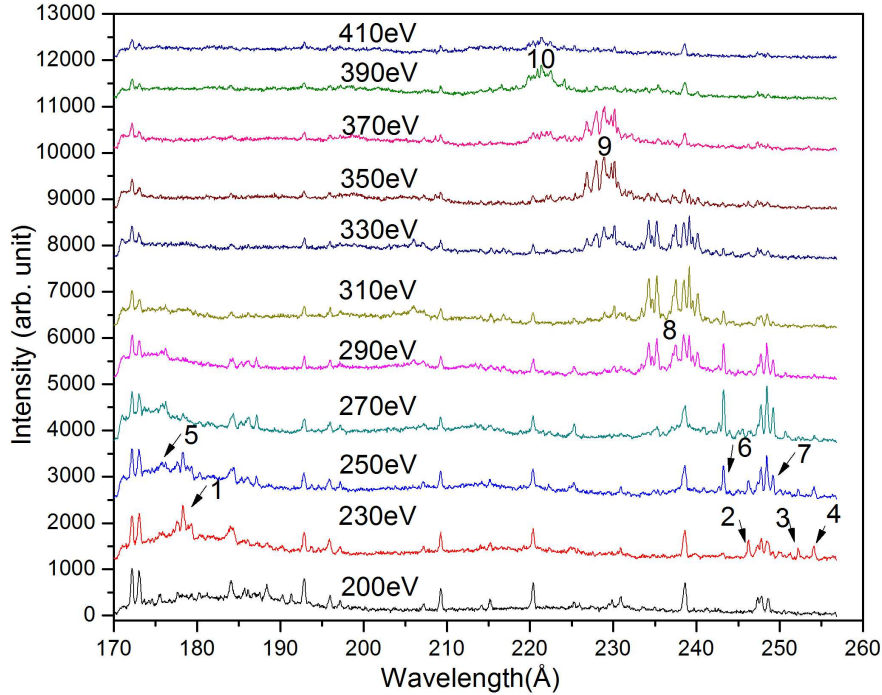}
\vspace{-0.4cm}
\caption{\label{exp} Spectra taken at SH-HtscEBIT at 11 electron beam energies from 200 to 400 eV, with a beam current of 8.1 mA, while injecting $\mathrm{W(CO)_6}$. Based on the ionization potentials given in table~\ref{ionization}, the lines labeled 1, 2, 3 and 4 belong to $\mathrm{W^{11+}}$ lines, 5, 6 and 7 belong to $\mathrm{W^{12+}}$ lines. Transition array 8 is from $\mathrm{W^{13+}}$, 9 is from $\mathrm{W^{14+}}$ and 10 is from $\mathrm{W^{15+}}$.}
\end{figure*}
The measurements reported in this work scanned the wavelength region of 100~-~400 \AA{}, but all identified lines fall in the 160~-~270 \AA ~region. The spectral development as a function of electron beam energy, and therefore the charge state of the tungsten-ions, is shown in figure~\ref{exp}.
\begin{table*}
\caption{\label{ionization}The ionization potential of $\mathrm{W^{11+}}$ to $\mathrm{W^{15+}}$~\cite{Kramida2009}.}
\begin{tabular}{llccccccccccccccc}
\hline
\hline
charge state   &&   $\mathrm{W^{10+}}$   &&  $\mathrm{W^{11+}}$   &&   $\mathrm{W^{12+}}$   &&    $\mathrm{W^{13+}}$    &&   $\mathrm{W^{14+}}$   &&   $\mathrm{W^{15+}}$ \\
ionization energy~(eV)&& $208.9\pm1.2$  &&   $231.6\pm1.2$   &&   $258.2\pm1.2$  &&  $290.7\pm1.2$   &&  $325.3\pm1.5$  &&  $361.9\pm1.5$  \\
\hline
\hline
\end{tabular}
\end{table*}

\begin{figure*}
\centering
\includegraphics[width=0.85\textwidth]{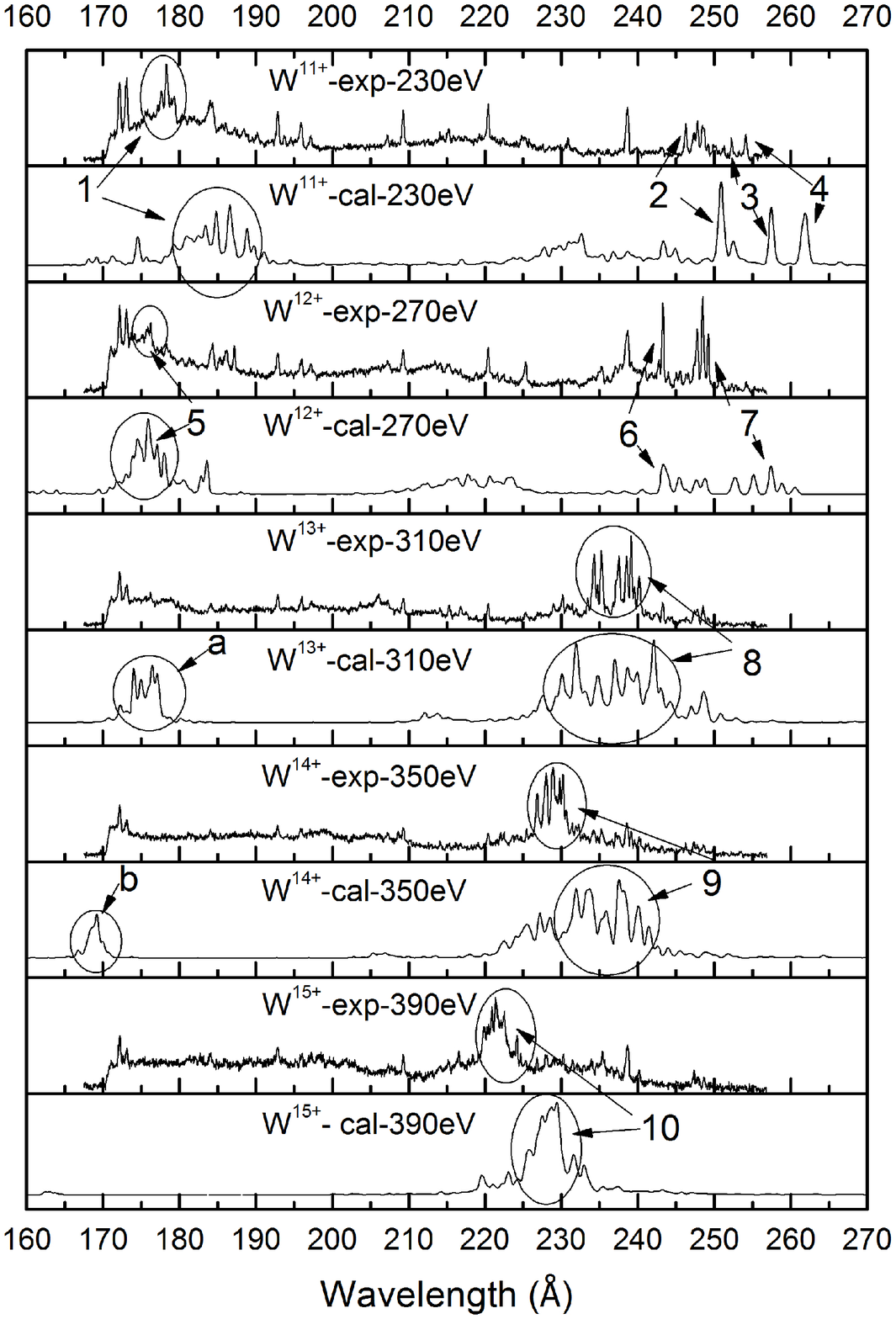}
\vspace{-0.4cm}
\caption{\label{exp&theo} Experimental (exp) and synthetic (cal) EBIT-spectra of $\mathrm{W^{11+}}$ to $\mathrm{W^{15+}}$ at electron energies of 230, 270, 310, 350 and 390 eV, respectively. The calculated lines were convoluted with a line width to resemble an instrument resolution of 850. The numbers refer to the same transitions as in figure~\ref{exp} and table~\ref{WL}. Lines labeled ``a" and ``b" in the synthetic spectra are not possible to observe in the experiment due to the Al foil transmission properties (giving rise to the cut-off around 170 {\AA}).}
\end{figure*}
Based on the ionization potentials given in table~\ref{ionization}, we assign the features labelled 1, 2, 3 and 4, which appear at electron beam energies between 200 and 230 eV, to the spectrum of $\mathrm{W^{11+}}$. In the same way, we conclude that the features labelled 5, 6 and 7 belong to $\mathrm{W^{12+}}$ lines, the transition array 8 is from $\mathrm{W^{13+}}$, 9 is from $\mathrm{W^{14+}}$ and 10 is from $\mathrm{W^{15+}}$. The identification of the $\mathrm{W^{13+}}$ arrays are further supported by the fact that they appear at the same beam energy as the visible lines in the same ion, as identified in \cite{Zhao2014}. In figure~\ref{exp&theo} we compare these experimental spectra with our synthetic ones from the CR-model spectra at some selected electron energies, where a single charge state is dominating for each case. We also list our computed ground states of $\mathrm{W^{11+}}$ to $\mathrm{W^{15+}}$ which agrees wth earlier results by Kramida and Shirai~\cite{Kramida2009}.

The synthetic spectra give us important guidance when identifying the experimental results, but there are two clear and systematic deviations. First, there is a shift towards longer wavelengths in the calculations, and second the transition arrays are too wide in the synthetic spectra. It may be possible to correct for these differences by including more correlation, such as core-core (CC) correlation~\cite{Fei2012}, but since we are aiming for CR-modelling with a large number of atomic properties, we have to limit the size of our atomic structure model. It is clear, though, that the features of the synthetic spectra can be used to identify the transition arrays appearing for each of the tungsten charge states.

The lines  are for all ions concentrated in the two wavelength regions (see figure~\ref{exp&theo}) 160~-~190~{\AA}~and 210~-~260~{\AA}. The former corresponds to $5d\rightarrow 5p$ transitions, while the latter arises from $5p\rightarrow 5s$ arrays. The results are in agreement with Suzuki's work~\cite{Suzuki2011}, in which their calculation showed $5p\rightarrow5s$ transitions dominating the spectrum around 250~\AA~for the tungsten charge states of W$^{12+}$-W$^{15+}$.
Our detailed identifications are listed in table~\ref{WL} and we will discuss them for each ion in detail in the following paragraphs.
\begin{table*}
\caption{\label{WL}Identified lines and arrays in $\mathrm{W^{11+}}~-~\mathrm{W^{15+}}$. The labels in the second column refer to figure~\ref{exp&theo}. The wavelengths are in unit of {\AA}. The uncertainties of the experimental wavelengths in transition arrays are around 0.15~\AA~and mainly due to the statistical uncertainty of the wavelength calibration and fitting.}
\footnotesize
\begin{tabular}{cccccccllllllllll}
\hline
\hline
ion    & label    &  experimental   &&    theoretical && transitions    && type       \\
    &    &   wavelength  &&  wavelength &&            \\
\hline
              &    1   &    175.6~-~180.4     &&  183.3~-~189.0     &&  $4f^{13}5s^25p5d$   $\rightarrow$   $4f^{13}5s^25p^2$ && array \\
$\mathrm{W^{11+}}$     &    2   &     246.23~$\pm$~0.12     &&  250.84     &&  $[[(4f^{13})_{5/2}5s]_3(5p^2_{1/2}5p_{3/2})_{3/2}]_{3/2}$   $\rightarrow$   $[[(4f^{13})_{5/2}5s^2]_{5/2}(5p^2_{1/2})_0]_{5/2}$ && line   \\
              &    3   &     252.25~$\pm$~0.18     &&  257.46     &&  $[[(4f^{13})_{5/2}5s]_2(5p^2_{1/2}5p_{3/2})_{3/2}]_{7/2}$   $\rightarrow$   $[[(4f^{13})_{5/2}5s^2]_{5/2}(5p^2_{1/2})_0]_{5/2}$ && line  \\
              &   4   &     254.09~$\pm$~0.17     &&  261.98     &&  $[[(4f^{13})_{5/2}5s]_3(5p^2_{1/2}5p_{3/2})_{3/2}]_{5/2}$   $\rightarrow$  $[[(4f^{13})_{5/2}5s^2]_{5/2}(5p^2_{1/2})_0]_{5/2}$&& line    \\
\hline
              &    5   &    174.8~-~176.8  &&  173.8~-~178.0     &&  $4f^{12}5s^25p5d$    $\rightarrow$ $4f^{12}5s^25p^2$    && array    \\
$\mathrm{W^{12+}}$     &    6   &    243.27~$\pm$~0.13      &&   243.29    &&  $[[(4f^{13})_{7/2}5s]_45p_{1/2}]_{9/2}5p_{3/2}]_4$   $\rightarrow$ $[(4f^{13})_{7/2}5s^25p_{1/2}]_4$   && line     \\
              &    7   &    249.19~$\pm$~0.13      &&   257.43    &&  $[[(4f^{13})_{7/2}5s]_35p_{1/2}]_{7/2}5p_{3/2}]_5$   $\rightarrow$ $[(4f^{13})_{7/2}5s^25p_{1/2}]_4$     && line     \\
\hline
     &    a    &   &&  173.3~-~178.08  && $4f^{12}5s^25d$  $\rightarrow$  $4f^{12}5s^25p$ && array  \\
$\mathrm{W^{13+}}$              &    8   & 232.7~-~240.9  &&  225.8~-~245.9     &&  $\left \{
\begin{array} [c] {lllllllllll}
4f^{13}5s5p    \rightarrow   4f^{13}5s^2 \\
4f^{12}5s5p^2   \rightarrow  4f^{12}5s^25p
\end{array} \right.$ && array \\
\hline
     &    b   &  &&  166.6~-~171.3  &&   $4f^{11}5s^25d$  $\rightarrow$    $4f^{11}5s^25p$   && array\\
$\mathrm{W^{14+}}$              &    9   &  225.4~-~233.7  &&   221.5~-~242.7    &&    $\left \{
\begin{array} [c] {lllllllllll}
4f^{12}5s5p    \rightarrow    4f^{12}5s^2  \\
4f^{11}5s5p^2    \rightarrow    4f^{11}5s^25p
\end{array} \right.$&& array \\
\hline
$\mathrm{W^{15+}}$     &    10   &  214.8~-~226.9  &&   218.4~-~234.5    &&   $4f^{11}5s5p$ $\rightarrow$ $4f^{11}5s^2$     && array    \\
\hline
\hline
\end{tabular}
\end{table*}

\paragraph{$\mathrm{W^{11+}}$}{} Our predicted ground state of $\mathrm{W^{11+}}$ is $4f^{13}5s^25p^2$~($[[(4f^6_{5/2})_0(4f^7_{7/2})_{7/2}]_{7/2}5s^2(5p^2_{1/2})_0]_{7/2}$ in~jj-coupling). The experimental and synthetic spectra are shown in figure~\ref{exp&theo} at E~=~230eV. According to the spectra, four transition arrays or lines are determined to be from this ion. $4f^{13}5s^25p5d$ $\rightarrow$ $4f^{13}5s^25p^2$ transitions contribute to the 175.6~-~180.4~\AA{} array(labeled as ``1" in figure~\ref{exp&theo}) and the three longer-wavelength lines we identify as $4f^{13}5s5p^3$ $\rightarrow$ $4f^{13}5s^25p^2$ transitions.

\paragraph{$\mathrm{W^{12+}}$}{} In the present work the ground state of $\mathrm{W^{12+}}$ is $[4f^{14}5s^2]_0$. We also compared our computational results with available data in \cite{Safronova2013a}. The differences between our RCI results and the various results from \cite{Safronova2013a}, i.e. RMBPT1(First-order relativistic many-body perturbation theory), RMBPT2(Second-order relativistic many-body perturbation theory) and COWAN(Hartree-Fock relativistic method) values of excitation energies are 1-5\%, 0-3\% and 1-4\%, respectively. Moreover, our results are in better agreement with the RMBPT2 results and most of the excitation energies agree within 2\%. We would like to point out that in spite of the fact that the ground state suggested from our calculations is the same as the ones given by Kramida and Safronova's works~\cite{Kramida2009,Safronova2013a}, this is only a tentative identification. The level splitting of the  $4f^{12}5s^25p^2$ configuration is very sensitive to the computational model, and the inclusion of more correlation migth push the lowest level in this configuration below the $4f^{14}5s^2~\mathrm{^1S_0}$ level. A larger calculations will be needed to resolve this issue. The experimental and synthetic spectra are shown in figure~\ref{exp&theo} for an energy of E~=~270eV. The transition array ``5" in the wavelength region of 174.8~-~176.8~\AA{} in the experimental spectra(see figure~\ref{exp&theo}) is identified to belong to the $4f^{12}5s^25p5d$ $\rightarrow$ $4f^{12}5s^25p^2$ array, while the other two lines at 243.3~\AA{} and 249.2~\AA{} are $5p$ $\rightarrow$ $5s$ transitions, outside a $4f^{13}_{7/2}$-subshell.

\begin{figure}
\centering
\includegraphics[width=0.85\textwidth]{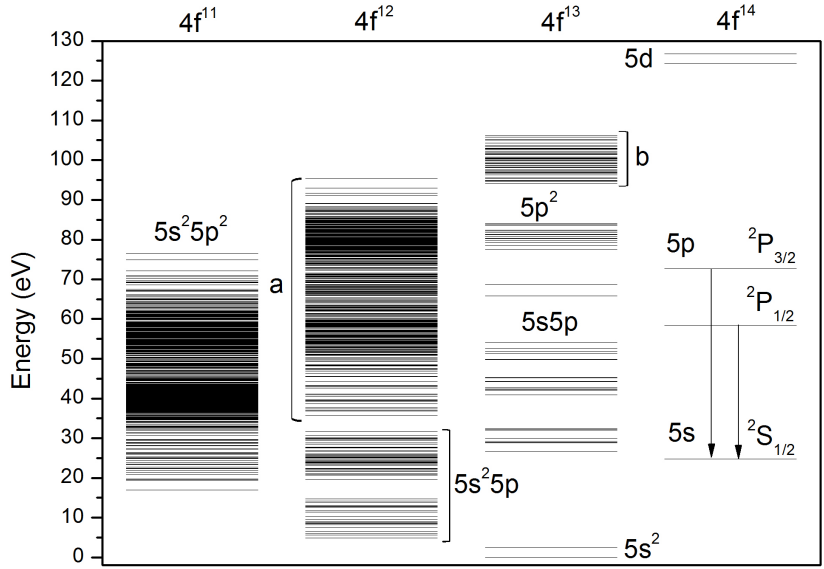}
\vspace{-0.4cm}
\caption{\label{structure3} Energy level structure of $\mathrm{W^{13+}}$. ``a" refers to energy levels of the $4f^{12}5s^25d$ and $4f^{12}5s5p^2$ configurations, while ``b" gives energy levels of the $4f^{13}5s5d$ and $4f^{13}5p^2$. Lines of $\mathrm{W^{13+}}$ observed in this work are mainly from $4f^{12}5s^25d$  $\rightarrow$  $4f^{12}5s^25p$, $4f^{13}5s5p$    $\rightarrow$  $4f^{13}5s^2$ and $4f^{12}5s5p^2$   $\rightarrow$  $4f^{12}5s^25p$ transitions. The ``Curtis and Ellis" lines (see text) are marked in the right portion of the figure.}
\end{figure}
\paragraph{$\mathrm{W^{13+}}$}{} The experimental and synthetic spectra of this Promethium-like ion are shown in figure~\ref{exp&theo} at E~=~310eV. There is still some remaining controversy concerning the ground-state of $\mathrm{W^{13+}}$ because of the effect of the ``collapse" of the $4f$-orbital, which will lead to a change in ground-state configuration of promethium-like ions from $4f^{13}5s^2$ to $4f^{14}5s$, leading to a simpler semi-one-electron spectrum, with two strong resonance lines between the $5s$ and $5p$ levels (see figure~\ref{structure3}). In 1980, Curtis and Ellis predicted that this change would happen for $Z < 74$, by using the Hartree-Fork method~\cite{Curtis1980}. Later, in 1983, Theodosiou indicated that the change occurred first at Z~=~78 by using the Dirac-Fork method~\cite{Theodosiou1983}. Recent calculations~\cite{Kramida2006,Vilkas2008,Safronova2013} support the prediction of Theodosiou. In 2003, Hutton et al. reported a detection of the 5p $\rightarrow$ 5s resonance lines in a spectrum from the Berlin-EBIT~\cite{Hutton2003}, which would support the model by Curtis and Ellis, and contradicting later theoretical predictions. However, the performance of this high-energy EBIT, working at only a few hundreds of eV energy, is quite debatable and leads to an uncertain and tentative identification.
\begin{table*}
\caption{\label{W13resonance} Wavelength for the $4f^{14}5p~\mathrm{^2P_{1/2,3/2}}\rightarrow4f^{14}5s~\mathrm{^2S_{1/2}}$ transitions in $\mathrm{W^{13+}}$. The values are in unit of \AA.}
\begin{tabular}{llccccccccccccccc}
\hline
\hline
transitions && RCI  && RMBPT$\mathrm{^a}$ && RMBPTB$\mathrm{^b}$ && COWAN$\mathrm{^c}$ && MR-MP$\mathrm{^d}$ &&   exp$\mathrm{^e}$ \\
\hline
$\mathrm{^2P_{1/2}}\rightarrow\mathrm{^2S_{1/2}}$ && 367.52 && 370.92 && 371.04 &&  367.95 && 369.14  &&  365.3 \\
$\mathrm{^2P_{3/2}}\rightarrow\mathrm{^2S_{1/2}}$ && 257.83 &&260.94  && 261.12 && 260.83 && 259.62 &&  258.2 \\
\hline
\hline
\end{tabular}
\footnotetext[1]{Second-order approximation with a Dirac-Fock (DF) potentials \cite{Safronova2013}.}
\footnotetext[2]{Second-order approximation with a Breit-Dirac-Fock (BDF) potentials \cite{Safronova2013}.}
\footnotetext[3]{Hartree-Fock relativistic method implemented in the COWAN code \cite{Safronova2013}.}
\footnotetext[4]{Multireference M$\phi$ller-Plesset~(MR-MP) approach \cite{Vilkas2008}.}
\footnotetext[5]{Experimental wavelengths from Ref.~\cite{Hutton2003}.}
\end{table*}

The wavelengths of the $4f^{14}5p~\mathrm{^2P_{1/2}}\rightarrow4f^{14}5s~\mathrm{^2S_{1/2}}$ and $4f^{14}5p~\mathrm{^2P_{3/2}}\rightarrow4f^{14}5s~\mathrm{^2S_{1/2}}$ transitions for $\mathrm{W^{13+}}$ are listed in table \ref{W13resonance}. In the present work our calculations predict the $4f^{14}5p~\mathrm{^2P_{1/2,3/2}}\rightarrow4f^{14}5s~\mathrm{^2S_{1/2}}$ transition lines at 367.52 \AA~and 257.83 \AA. The biggest difference between our RCI and other theoretical results for the $\mathrm{^2P_{1/2}}\rightarrow\mathrm{^2S_{1/2}}$ and $\mathrm{^2P_{3/2}}\rightarrow\mathrm{^2S_{1/2}}$ transitions are 0.96\% and 1.27\%, respectively. The difference between our RCI and the experimental results for the $\mathrm{^2P_{1/2}}\rightarrow\mathrm{^2S_{1/2}}$ and $\mathrm{^2P_{3/2}}\rightarrow\mathrm{^2S_{1/2}}$ transitions are 0.14\% and 0.60\%, respectively.
The ground state configuration is predicted to be $4f^{13}5s^2$ and our CR-model do not indicate any strong $4f^{14}5p \rightarrow 4f^{14}5s$ lines. This is in agreement with our experiment where we did not observe any strong resonance line around the predicted position (see table~\ref{W13resonance}) with a simple structure. Figure~\ref{structure3} shows the level structure of $\mathrm{W^{13+}}$. Clearly there are numerous levels lower in energy than the $4f^{14}5s$ and $4f^{14}5p$. This leads to a low population of the $4f^{14}5p$~levels and hence weak lines as opposed to the strong lines predicted by Curtis and Ellis~\cite{Curtis1980}. This situation is similar to $\mathrm{W^{11+}}$ and $\mathrm{W^{12+}}$, where the simplest configurations, with closed $4f^{14}$-subshell configurations, are not the ground state. The occurrence of simple spectra, with few resonance lines, will therefore occur for higher Z in all the corresponding iso-electronic sequences.
To further support the identification of $4f^{13}5s^2$ as the ground-state configuration of $\mathrm{W^{13+}}$, we recently observed visible M1 transitions within the $4f^{13}5s^2$ and $4f^{12}5s^25p$ configurations using the SH-HtscEBIT~\cite{Zhao2014}. The transition array ``8" in figure~\ref{exp&theo} observed at around 232.7~-~240.9~\AA{} was identified as $5p$ $\rightarrow$ $5s$ arrays outside a $4f^{13}$ or $4f^{12}$ subshell (see table~\ref{WL}), which coincide with the results of Safronova~\cite{Safronova2013}. A second strong transition array ``a" in figure~\ref{exp&theo}, positioned at 173.3~-~178.08~\AA{}  arises from $5d$ $\rightarrow$ $5p$ transitions, outside a $4f^{12}$ subshell, but is unobservable in the experimental spectrum, due to the wavelength cut-off of the Aluminium foil.

\paragraph{$\mathrm{W^{14+}}$}{} We predict the ground state of $\mathrm{W^{14+}}$ to be $4f^{12}5s^2$~($[[(4f^6_{5/2})_0(4f^6_{7/2})_6]_65s^2]_6$ in~jj-coupling). The experimental and synthetic spectra are shown in figure~\ref{exp&theo} at E~=~350eV. The transition array ``9" in figure~\ref{exp&theo} covering the wavelength region of 225.4~-~233.7~\AA{} is suggested to be from $5p$ $\rightarrow$ $5s$ transitions outside a $4f^{12}$ or $4f^{11}$ subshell (see table~\ref{WL}). The shorter wavelengths ``b" in figure~\ref{exp&theo} at around the 166.6~-~171.3~\AA{} wavelength region arises from $4f^{11}5s^25d$  $\rightarrow$    $4f^{11}5s^25p$ transitions. Again these lines did not appear in the experimental spectrum due to the Al foil transmission properties.

\paragraph{$\mathrm{W^{15+}}$}{} Our predicted ground state of $\mathrm{W^{15+}}$ is $4f^{11}5s^2$~($[[(4f^6_{5/2})_0(4f^5_{7/2})_{15/2}]_{15/2}5s^2]_{15/2}$ in jj-coupling). The experimental and synthetic spectra are shown in figure~\ref{exp&theo} at E~=~390eV. The transition array ``10" at 214.8~-~226.9~\AA{} in figure~\ref{exp&theo} are identified as from $4f^{11}5s5p$ $\rightarrow$ $4f^{11}5s^2$.
\section{Conclusion}
To conclude, we have studied EUV spectra from $\mathrm{W^{11+}}$ to $\mathrm{W^{15+}}$ ions experimentally and theoretically. The experiments were carried out using a new flat field spectrometer installed at the SH-HtscEBIT. The spectra were recorded at electron beam energies ranging between 200 and 400 eV and spectral features were recorded in the 170~-~260~\AA~region. In order to identify the charge states and spectral features we performed detailed RCI calculations of the relevant atomic structure. The data was then used in a CR-model of the EBIT. Both calculations are a part of the FAC code. Although there were systematic shifts in the calculated wavelengths and the widths of the spectral features were a systematically wider than the corresponding experimental arrays, the synthetic spectra are in good enough overall agreement with experiment, to facilitate a number of identifications of arrays and lines. The spectral features in the 160~-~190~\AA{} region are from $5d$ $\rightarrow$ $5p$ transitions whereas $5p$ $\rightarrow$ $5s$ falls in the 210~-~260~\AA{} region, for all the ions considered. This work also identified the ground-state configuration of $\mathrm{W^{13+}}$ to be $4f^{13}5s^2$, supported by both experimental spectra and theoretical calculations.

\section{Acknowledgements}
This work was supported by the Chinese National Fusion Project for ITER No. 2015GB117000 and Shanghai Leading Academic Discipline Project No. B107. WL is grateful for support from the Nordic Center at Fudan University for the exchange of staff between Fudan and Lund University.

%end{CJK*}

\section*{References}
\begin{thebibliography}{99}

\bibitem[1]{Putterich2005}
T P\"utterich, R Neu, C Biedermann, R Radtke, and A U Team,
\newblock {J. Phys. B: At. Mol. Opt. Phys.} \textbf{38}, 3071 (2005).

\bibitem[2]{Skinner2008}
C H Skinner,
\newblock {Can. J. Phys.} \textbf{86}, 285 (2008).

\bibitem[3]{Philipps2006}
V Philipps,
\newblock {Phys. Scr.} \textbf{T123}, 24 (2006).

\bibitem[4]{Robin2014}
\newblock {R Barnsley, private communication and
https://www-amdis.iaea.org/meetings/AMPMI14/Presentations \\
/AMPMI-2014-12-15-Talk-Barnsley-ITER-2by4.pdf, the Decennial IAEA Technical Meeting on Atomic, Molecular and Plasma-Material Interaction Data for Fusion Science and Technology, Daejeon, South Korea, December 2014.
}

\bibitem[5]{NIST}
\newblock {NIST Atomic Spectra Database, Version 5}: http://www.nist.gov/pml/data/asd.cfm

\bibitem[6]{Trabert1986}
E Tr\"abert, and P Heekmann
\newblock {Z. Phys. D} \textbf{1(4)}, 381¨C383 (1986).

\bibitem[7]{Kramida2011}
A Kramida,
\newblock {Can. J. Phys.} \textbf{89}, 551¨C570 (2011).

\bibitem[8]{Zhao2014}
Z Zhao \emph{et al.},
\newblock {J. Phys. B: At., Mol. Opt. Phys.} accepted (2014).

\bibitem[9]{Li2015}
Li \emph{et al.}, in preparation, http://arxiv.org/abs/1503.04523

\bibitem[10]{Fei2014}
Z Fei, W Li, J Grumer, Z Shi, R Zhao, T Brage, S Huldt, K Yao, R Hutton, and Y Zou,
\newblock {Phys. Rev. A} \textbf{90}, 052517 (2014).

\bibitem[11]{Fei2012}
Z Fei, R Zhao, Z Shi, J Xiao, M Qiu, J Grumer, M Andersson, T Brage, R Hutton, and Y Zou,
\newblock {Phys. Rev. A} \textbf{86}, 062501 (2012).

\bibitem[12]{ruifeng2014}
R Zhao, J Grumer, W Li, J Xiao, S Huldt, T Brage, R Hutton, and Y Zou,
\newblock {J. Phys. B: At., Mol. Opt. Phys.} \textbf{47}, 185004 (2014).

\bibitem[13]{Grumer2014}
J Grumer, R Zhao, T Brage, S Huldt, R Hutton, and Y Zou,
\newblock {Phys. Rev. A}, \textbf{89}, 062511 (2014)

\bibitem[14]{Qiu2014}
M Qiu \emph{et al.},
\newblock {J. Phys. B: At., Mol. Opt. Phys.} \textbf{47}, 175002 (2014).

\bibitem[15]{Zhan2014}
Z Shi, R Zhao, W Li, B Tu, Y Yang, J Xiao, S Huldt, R Hutton, and Y Zou,
\newblock {Rev. Sci. Instrum.} \textbf{85}, 063110 (2014).

\bibitem[16]{Gu2008}
M F Gu,
\newblock {Can. J. Phys.} \textbf{86}, 675-689 (2008).

\bibitem[17]{Curtis1980}
L Curtis, and D Ellis,
\newblock {Phys. Rev. Lett.} \textbf{45(26)}, 2099 (1980).

\bibitem[18]{Xiao2014}
J Xiao, R Zhao, X Jin, B Tu, Y Yang, D Lu, R Hutton, and Y Zou,
\newblock {Proc. IPAC2013} MOPFI066 (2013).

\bibitem[19]{Kita1983}
T Kita, T Harada, N Nakano, and H Kuroda,
\newblock {Appl. Opt.} \textbf{22}, 512 (1983).

\bibitem[20]{Harada1980}
T Harada, and T Kita,
\newblock {Appl. Opt.} \textbf{19}, 3987 (1980).

\bibitem[21]{Harada1999}
T Harada, K Takahashi, H Sakuma, and A Osyczka,
\newblock {Appl. Opt.} \textbf{38}, 2743 (1999).

\bibitem[22]{Ken2000}
K K Lai, and H H Lamb,
\newblock {Thin Solid Films} \textbf{370}, 114-121 (2000).

\bibitem[23]{Edlen1963}
B Edl$\acute{e}$n,
\newblock {Rep. Prog. Phys.} \textbf{26}, 181-212 (1983).

\bibitem[24]{Bockasten1968}
K Bockasten, and K B Johansson,
\newblock {Ark. Fys.} \textbf{38}, 563-584 (1968).

\bibitem[25]{Moore1971}
C E Moore,
\newblock {Nat. Stand. Ref. Data Ser.} \textbf{4}, 46 (1971).

\bibitem[26]{Edlen1934}
B Edl$\acute{e}$n,
\newblock {Nova Acta Reg. Soc. Sci. Upsalien., Ser. } \textbf{9(6)}, 1-153 (1934).

\bibitem[27]{Gu2004}
M F Gu,
\newblock {Phys. Rev. A} \textbf{70}, 062704 (2004).

\bibitem[28]{Gu2005}
M F Gu,
\newblock {Phys. Rev. A} \textbf{156}, 105 (2005).

\bibitem[29]{Gu2006}
M F Gu, T Holczer, E Behar, and S M Kahn,
\newblock {Phys. Rev. A} \textbf{641}, 1227 (2006).

\bibitem[30]{Ralchenko2007}
Y Ralchenko,
\newblock {J. Phys. B: At. Mol. Opt. Phys.} \textbf{40}, F175 (2007).

\bibitem[31]{Kramida2009}
A Kramida, and T Shirai,
\newblock {At. Data Nucl. Data Tables} \textbf{95}, 305-474 (2009).

\bibitem[32]{Suzuki2011}
C Suzuki, C S Harte, D Kilbane, T Kato, H A Sakaue, I Murakami, D Kato, K Sato, N Tamura, S Sudo, M Goto, R D¡¯Arcy, E Sokell and G O¡¯Sullivan
\newblock {J. Phys. B: At. Mol. Opt. Phys.} \textbf{44}, 175004 (2011).

\bibitem[33]{Safronova2013a}
U I Safronova, A S Safronova, and P Beiersdorfer,
\newblock {Phys. Rev. A} \textbf{87}, 032508 (2013).

\bibitem[34]{Theodosiou1983}
C E Theodosiou, and V Raftopoulos,
\newblock {Phys. Rev. A} \textbf{28(2)}, 1186-1188 (1983).

\bibitem[35]{Kramida2006}
A Kramida, and J Reader,
\newblock {At. Data Nucl. Data Tables} \textbf{92(4)}, 457-479 (2006).

\bibitem[36]{Vilkas2008}
M J Vilkas, Y Ishikawa, and E Tr\"abert,
\newblock {Phys. Rev. A} \textbf{77}, 042510 (2008).

\bibitem[37]{Safronova2013}
U I Safronova, A S Safronova, and P Beiersdorfer,
\newblock {Phys. Rev. A} \textbf{88}, 032512 (2013).

\bibitem[38]{Hutton2003}
R Hutton, Y Zou, J R Almandos, C Biedermann, R Radtke, A Greier, and R Neu,
\newblock {Nucl. Instrum. Methods Phys. Res., Sect. B} \textbf{205}, 114-118 (2003).

\end{thebibliography}
\end{document}